\documentclass[%
reprint,
amsmath,amssymb,
aps,
]{revtex4-1}
\usepackage{amsmath}
\usepackage{empheq}
\usepackage{graphicx}% Include figure files
\usepackage{dcolumn}% Align table columns on decimal point
\usepackage{bm}% bold math
\begin{document}
	
	\preprint{APS/123-QED}
	
	\title{Particle Creation in a Linear Gravitational Wave Background}% Force line breaks with \\

	\author{Tanmoy Chakraborty} 	
	\email{ug19tc2135@iacs.res.in}
	
	\author{Parthasarathi Majumdar}
	\email{parthasarathi.majumdar@iacs.res.in}
	
	\affiliation{%
		Indian Association for the Cultivation of Science\\Raja SC Mallick Road, Jadavpur, Kolkata 700032, India 
	}%

	\date{05 February 2024}% It is always \today, today,
	
	\begin{abstract}

Inspired by Leonard Parker's pioneering 1968 work demonstrating matter quanta production in a dynamical spacetime background, we consider production of scalar quanta in a gravitational wave background. Choosing the spacetime to be a flat spacetime perturbed linearly by a linear gravitational wave, we show that scalar particles may indeed be produced in a perturbative manner. Our formulation is valid for any linear gravitational wave background profile, and is by no means restricted to monochromatic plane waves, in contrast to much of the earlier work on this topic. Thus, our work is directly applicable to gravitational wave signals from compact binary coalescence detected at LIGO-VIRGO-KAGRA, where they are of a pulsed character rather than monochromatic plane waves. We also briefly outline generalizing our approach for photon creation in a gravitational wave background. In this aspect, irrespective of the astrophysical nature of the binary merger sourcing the gravitational wave signal, one expects the dynamical nature of the spacetime to produce all species of light particles. Thus, any binary coalescence is in effect a source of multimessenger astrophysics.          	
		
\end{abstract}
	
	\maketitle
	
	%\tableofcontents
	
	\section{Introduction}
	
The vacuum state $|0\rangle$ in quantum field theory in a flat spacetime background  is characterized by the generators of the Poincare algebra, annihilating it, as follows: 
\begin{align*}
\mathcal{P}^{\mu}|0\rangle&=0 \\ \mathcal{M}^{\mu\nu}|0\rangle&=0
\end{align*}
 where $\mathcal{P}^{\mu}$ are generators of spacetime translations and $\mathcal{M}^{\mu\nu}$ are the generators of Lorentz boosts and rotations. The commutation relations which form the algebra hold true only when the spacetime is flat, rendering the vacuum a unique state in the multiparticle Hilbert space. This of course is no longer true in a general curved spacetime. However, because of the local Lorentz invariance of general relativity, at every point (event) there is a frame in which such a vacuum may exist. This implies that the vacuum state changes from point to point in a curved background, and so the global uniqueness of the vacuum state is lost. 
 
If we consider a curved spacetime which is asymptotically flat at early times $(t\rightarrow-\infty)$ and at late times $(t\rightarrow+\infty)$, we can define the vacuum state at early times $|0\rangle_-$ and the vacuum state at late times $|0\rangle_+$ with no ambiguity whatsoever. This definition of vacuum state only holds asymptotically, because the spacetime in between might be curved in general. Hence in general, $|0\rangle_-\neq|0\rangle_+$. Any quantum field can be quantised using the usual canonical commutation rules, at the asymptotic regions where the spacetime is flat. The particle states are then defined as excitations around $|0\rangle_\pm$. As the vacuum state has no particles in it, the number operator corresponding to that vacuum state when acted upon the state should vanish. So in the context of our early time and late time vacuum states, we must have, $N_\pm|0\rangle_\pm=0$ where $N_\pm$ are the late and early time number operators, respectively. However, it is clear that $N_{\mp} |0\rangle_{\pm} \neq 0$ necessarily. This fact was first noticed by Leonard Parker \cite{par68} in a cosmological context, where the dynamical nature of the background curved spacetime led to production of quanta in any standard field theory. For the case of massive stars collapsing to a black hole, Hawking \cite{haw75} showed that $N_+|0\rangle_-\neq 0$ for a massless scalar field propagating in the dynamical background of the collapsing star. Or in other words, ${}_-\langle0|N_+|0\rangle_-\neq0$. Later, Parker \cite{par76} also showed this process of particle creation in case of an expanding universe, taking the FLRW spacetime. This means that there is a flux of created particles with respect to the early time vacuum, resulting from the changing curvature of the intermediate spacetime that connects the two asymptotic regions. More recently, Parker and Toms \cite{partom09} also showed that this phenomenon of particle creation is possible in case of any dynamical spacetime, which is a spacetime having no timelike isometries. 

Historically, Deser \cite{des75} and Gibbons \cite{gib75} had established the impossibility of creation of massive scalar quanta and photons in {\it plane} monochromatic gravitational waves. Later assays have tried to evade this `no-go' result by a variety of alternatives, like considering specific boundary conditions on the quantum scalar fields \cite{sor00}, considering {\it massless} instead of massive quantum matter fields \cite{sing16} and so on, but invariably in {\it plane} gravitational wave backgrounds. Thus, while massless scalar quanta and photon creation have been in principle demonstrated in the papers \cite{sor00} - \cite{sing23}, the restriction to plane gravitational waves in these papers implies that the more realistic situation of gravitational waves as pulsed signals in time domain, corresponding to the signals detected at the laser interferometer observatories since 2015, have not been considered in previous works. Our long-term aim is to be able to characterize every compact binary coalescence producing gravitational wave signals as an event in {\it multimessenger astronomy} through matter quanta production. In this pursuit, there is not much overlap of our work with earlier literature, with the value-addition of greater relevance to recent observational data. 

In this paper, we extend Parker's pioneering perspective to consider possible creation of matter quanta in a gravitational wave background, which of course is dynamical in the sense alluded to above. Recent direct observation of gravitational waves by LIGO and VIRGO and KAGRA (LVK), attibuted to compact binary coalescences, has led to classification of such coalescences as to whether they entail gravitational wave emission alongwith emision of electromagnetic waves or not. Thus, GW150914 is likely a result of binary black hole merger and hence classically involves only gravitational wave emission, while GW170817, which is a binary neutron star merger, involves short gamma ray emissions in addition to gravitational waves \cite{bnsmerg}. The latter is taken to signify the beginning of `multimessenger astrophysics'. On the other hand, if we include in this perspective production of matter field quanta of diverse species, in compact binary coalescence, then even a black hole merger may be interpreted as a multimessenger astrophysical event, once fluxes of emitted quanta are detected. Actual detection of such quanta may require highly sensitive instrumentation beyond the reach of current terrestrial capabilities. However, close to the merger event in space and time, a detectably large blast of emitted quanta cannot be ruled out. Here, we present a basic framework for emission of scalar quanta in interaction with an ambient linear gravitational wave, treated as a classical background. We cast the formulation in terms of the Fourier transform of the gravitational wave signal, so that the frequency spectrum of gravitational wave signals detected at LVK may serve as an input to our final result for massless scalar quanta production. Accordingly, our approach is by no means restricted to plane gravitational waves, but accommodates {\it all} gravitational wave signals which admit a Fourier spectrum. We also note that the produced quanta may have an anisotropic spatial distribution, which may also lead to detection strategies for these quanta. It is shown that the entire approach may be straightforwardly generalized to the case of photon creation in such a linear gravitational wave background with any well-defined nontrivial Fourier spectrum, and the corresponding Bogoliubov coefficients are derived.  

The paper is organized as follows : the next section (2) includes a description, in the weak-field approximation to general relativity, of the {\it transverse-traceless} gravitational wave in terms of projection operators replacing the standard gauge-fixing. We also mention that the amplitudes of these waves decay with spatial distance traversed, thereby satisfying asymptotic flatness of the full spacetime. In section 3, we consider the perturbative solution to the scalar field equations in this gravitational wave background. In section 4 we invoke the semiclassical approximation, introduce the Bogoliubov transformations relating the mode functions at early and late times and set up the formalism for computing the expectation value of the late time number operator in the early time vacuum, following ref. \cite{partom09}. This is used to present our main result for the spectrum of quanta produced, as a functional of the Fourer transform of the  gravitational wave background. Next we briefly outline generalizing our approach for photon creation in a gravitational wave background. We end in section 5 with a few concluding remarks.   

\section{Weak Field Approximation and Gauge Invariant Projections}
	
First we need to construct the metric for a gravitational wave propagating in a particular direction (say, x). So its a vacuum solution of the Einstein field equations:
\begin{eqnarray}
G_{\mu\nu}=R_{\mu\nu}-\frac{1}{2}g_{\mu\nu}R=0 \label{ee}
\end{eqnarray}
Since we will be doing linearised gravity, we need to make the so called weak field approximation:$$g_{\mu\nu}=\eta_{\mu\nu}+h_{\mu\nu}$$ where $|h_{\mu\nu}|<<1 \forall \mu, \nu=0,1,2,3$. So we retain terms linear in $h_{\mu\nu}$ at every step. We then have,
\begin{eqnarray}
\Gamma^{\alpha}_{\mu\nu}\approx\frac{1}{2}\eta^{\alpha\sigma}\left(\partial_{\mu}h_{\nu\sigma}+\partial_{\nu}h_{\mu\sigma}-\partial_{\sigma}h_{\mu\nu}\right) \label{ling}
\end{eqnarray} 
Under an infinitesimal displacement $\chi$, the Lie derivative
\begin{eqnarray*}
\mathcal{L}_{\chi}g_{\mu\nu} &=& \nabla_{\mu}\chi_{\nu}+\nabla_{\nu}\chi_{\mu} \label{gj}\\ 
{\rm or~~} \delta_{\chi}h_{\mu\nu} &=& \partial_{\mu}\chi_{\nu}+\partial_{\nu}\chi_{\mu} \label{gjh}
\end{eqnarray*}
It can be checked that under such a transformation, the Einstein's equations (\ref{ee}) remain invariant, that is, $$\delta_{\chi}G_{\mu\nu}=0$$ which implies that all components of $h_{\mu\nu}$ are not physical.

The standard approach, explained in all textbooks is to go through `gauge fixing', i.e., to `fix' the $\chi$ so that only the gauge-invariant physical $h_{\mu \nu}$ remain. We pursue an alternative approach, due to Anarya Ray \cite{aray19}, where one can {\it project out} the physical $h_{\mu \nu}$ degrees of freedom from the entire set. One starts with the projection operator ${\cal P}^{\mu}_{\nu}$ in Fourier space labelled by the 4-vector $k^{\mu}$ in Minkowski space classical electrodynamics \cite{pmar19} ; this is given by 
\begin{eqnarray}
{\cal P}^{\mu}_{\nu} & \equiv & \delta^{\mu}_{\nu} - \frac{k^{\mu} k_{\nu}} {k^2} ~,~ J_{\alpha} \neq 0  \label{Tn0}\\
& \equiv & \delta^{\mu}_{\nu} - k_{(+)}^{\mu} k_{(-)\nu} - k_{(-)}^{\mu} k_{(+) \nu} ~,~{\rm vacuum} ~\label{vac}
\end{eqnarray}
This satisfies the properties
\begin{eqnarray}
{\cal P}^{\mu}_{\nu} {\cal P}^{\nu}_{\rho} &=& {\cal P}^{\mu}_{\rho} \label{proj} \\
k^{\mu} {\cal P}_{\mu \nu} &=& 0 = k^{\nu} {\cal P}_{\mu \nu} \label{tr} 
\end{eqnarray}
Observe that in vacuum electrodynamics, one has two linearly independent null vectors $k_{(\pm)}^{\mu}$ to implement this projection operator, with $k^{\mu}_{(+)} k_{(-) \mu} = 1$ as a normalization. The projection operator ${\cal P}^{\mu}_{\nu}$ projects out the physical, {\it gauge-invariant} part of $A^{\mu}$, viz., $A_P^{\mu}$
\begin{eqnarray}
A_P^{\mu} & \equiv & {\cal P}^{\mu}_{\nu} A^{\mu} \nonumber \\
~\Rightarrow~ \partial_{\mu} A^{\mu}_P &=& 0 \label{trns} \\
~\Rightarrow~ A_P^{(\omega) \mu} & \equiv & {\cal P}^{\mu}_{\nu} (A^{\nu} + \partial^{\nu} \omega) = A^{\mu}_P \label{ginv}
\end{eqnarray}
It is important to note that eqn. (\ref{trns}) is {\it not} a gauge-choice, despite appearances, it is a result out of a projection derived directly from the Maxwell equations without any arbitrary choices. It's existence is easily established from the fact that the Faraday field tensor $F_{\mu \nu}$ is completely dependent on this gauge-invariant projection and {\it not} on it's complement which is unphysical, which is why it is gauge invariant.

From the projection operator for electrodynamics, Ray \cite{aray19} has  constructed the projection operator $\Pi^{\mu \nu}_{\lambda \sigma}$ appropriate to linearized gravity, defined as, 
\begin{eqnarray}
\Pi^{\mu \nu}_{\lambda \sigma} \equiv {\cal P}^{\mu \nu} {\cal P}_{\lambda \sigma} - \frac{1}{2} ({\cal P}^{\mu}_{\lambda} {\cal P}^{\nu}_{\sigma} + {\cal P}^{\nu}_{\lambda} {\cal P}^{\mu}_{\sigma}) ~. \label{gpro} 
\end{eqnarray}
This projection operator projects out the gauge-invariant, physical components of the linearised metric fluctuation $h_{T \mu \nu}$ :
\begin{eqnarray}
h^{\mu \nu}_T & \equiv & \Pi^{\mu \nu}_{\lambda \sigma}~ h^{\lambda \sigma} \label{prj} \\
~\Rightarrow~ \partial_{\mu} h^{\mu \nu}_T &=& 0 \label{trnsv} \\
~{\rm and}~\Rightarrow~ \delta_{\chi} h_T^{\mu \nu}&=& 0~. \label{gjin}
\end{eqnarray}
It is straightforward to verify that this physical projection of the metric fluctuation satisfies the linearized Einstein equation with a conserved matter source $T^{\mu \nu}_m$,
\begin{eqnarray}
\Box h^{\mu \nu}_T = 8 \pi G T_m^{\mu \nu} \label{eet}
\end{eqnarray}
When $T^{\mu \nu}_m=0$, the above projection operator projects out the {\it transverse-traceless} components
\begin{eqnarray}
h^{\mu \nu}_{TT} & \equiv & \Pi^{\mu \nu}_{\lambda \sigma}~ h^{\lambda \sigma} \\
~\Rightarrow~ \partial_{\mu} h^{\mu \nu}_{TT} &=& 0 = h^{\mu}_{TT \mu} \label{tt} \\
\Box h^{\mu \nu}_{TT} &=& 0 \label{lee}
\end{eqnarray}

We restrict our focus on the transverse-traceless gravitational waves, but not restricting these waves to be plane waves. Rather, we impose the conditions
\begin{eqnarray}
\lim_{\substack{r\rightarrow\infty\\ t\rightarrow \pm \infty}}\bar{h}_{TT \mu\nu}\rightarrow0
\end{eqnarray}
so that spherical gravitational waves which vanish asymptotically are admissible. 
\section{The Scalar Field Equations}

In a general metric, the Klein-Gordon equation of motion is:
\begin{eqnarray}
\Box_g \Phi &=& 0 \nonumber \\
{\rm Or} ~\frac{1}{\sqrt{-g}}\partial_{\mu}\left(\sqrt{-g}g^{\mu\nu}\partial_{\nu}\Phi\right) &=& 0 \label{kge}
\end{eqnarray}
Weak gravity approximation dictates $g_{\mu\nu} \approx \eta_{\mu\nu}+h_{\mu\nu}$ where $|h_{\mu\nu}| << 1$. In the perturbative scheme adopted here, we also expand the scalar field perturbatively
\begin{eqnarray}
\Phi(x) &=& \Phi^{(0)}(x) + \Phi^{(1)}(x)   \nonumber \\
| \Phi^{(1)}(x) | & << & |\Phi^{(0)}(x)|. \label{ptsc}
\end{eqnarray}
The scalar field equations now decompose in the first order into
\begin{eqnarray}
\Box_{\eta}\Phi^{(0)} &=& 0 \label{phi0} \\
\Box_{\eta} \Phi^{(1)} &=& h^{\mu \nu}_{TT} \partial_{\mu} \partial_{\nu} \Phi^{(0)} ~.\label{phi1}
\end{eqnarray} 
	
\section{Particle creation}

We solve eqn.(\ref{phi0}) as 
\begin{equation}
\Phi^{(0)}(x) = \int\frac{d^4k}{(2\pi)^3}\delta(k^2)\left(f_ka_k+f_k^*a^\dagger_k\right) \theta(k^0) ~\label{p0}
\end{equation}
where the mode functions $f_k=e^{-ik.x}$. To solve eqn.(\ref{phi1}), we resort to Green's functions, 
\begin{eqnarray}
\Phi^{(1)}(x)=\int d^4x'\mathcal{G}(x-x')h^{\mu\nu}(x'){\partial_\mu}'{\partial_\nu}'\Phi^{(0)}(x')
\end{eqnarray}
and tranform to four dimensional Fourier space with appropriate convolution integrals. Rearranging the order of the integrals, performing the integral over $x'$ and one of the convolution integrals, we obtain
\begin{eqnarray}
&&\Phi^{(1)}(x) = -\int\frac{d^4k}{(2\pi)^3}\frac{e^{ik.x}}{k^2}\int\frac{d^4l}{(2\pi)^3}\delta(l^2)l_\mu l_\nu \cdot \nonumber \\
&&\cdot \left [a_l\tilde{h}^{\mu\nu}(k+l)\delta((k+l)^2) + a^\dagger_l\tilde{h}^{\mu\nu}(k-l)\delta((k-l)^2)\right]~. \label{p1}
\end{eqnarray}
Now we perform a shifting of the four-momentum $k\to(k-l)$ in the first term and $k\to(k+l)$ in the second term: 
\begin{eqnarray}
\Phi^{(1)}(x) & = &-\int\frac{d^4k}{(2\pi)^3}\delta(k^2)\int\frac{d^4l}{(2\pi)^3}\delta(l^2)l_\mu l_\nu\tilde{h}^{\mu\nu}(k) \cdot \nonumber \\
&& \left\{\frac{e^{i(k-l).x}}{(k-l)^2}a_l+\frac{e^{i(k+l).x}}{(k+l)^2}a_l^\dagger\right\}
\end{eqnarray}
Making use of the fact that $k^2=l^2=0$ from the delta functions and exchanging the dummy indices we have: 
\begin{eqnarray}
\Phi^{(1)}(x) & = &\frac{1}{2}\int\frac{d^4k}{(2\pi)^3}\delta(k^2)\int\frac{d^4l}{(2\pi)^3}\delta(l^2)k_{\mu}k_{\nu}\tilde{h}^{\mu\nu}(l)(k.l)^{-1} \cdot \nonumber \\
&& \cdot \left\{e^{-i(k-l).x}a_k-e^{i(k+l).x}a_k^\dagger\right\}
\end{eqnarray}

In the second integral, we let $l\to-l$ and get: 
\begin{eqnarray}
&& \Phi^{(1)}(x)= \frac{1}{2}\int\frac{d^4k}{(2\pi)^3}\delta(k^2)\int\frac{d^4l}{(2\pi)^3}\delta(l^2)k_{\mu}k_{\nu}(k.l)^{-1} \cdot \nonumber \\
&\cdot& \left\{\tilde{h}^{\mu\nu}(l)e^{-i(k-l).x}a_k+\tilde{h}^{\mu\nu}(-l)e^{i(k-l).x}a_k^\dagger\right\}
\end{eqnarray}

Thus we add $\Phi^{(0)}(x)$ and $\Phi^{(1)}(x)$ and obtain the complete perturbative solution in first order to the field equation : 
\begin{eqnarray}
&& \Phi(x)=\int\frac{d^4k}{(2\pi)^3}\delta(k^2) [e^{-ik.x} + \nonumber \\
&& +\frac{1}{2}\int\frac{d^4l}{(2\pi)^3}\delta(l^2)k_{\mu}k_{\nu} \{ \tilde{h}^{\mu\nu}(l)(k.l)^{-1}e^{-i(k-l).x}a_k \} \nonumber \\
&+& h.c.\} ]
\end{eqnarray}
where we have used the fact that $\tilde{h}^{\mu\nu}(-l)=\tilde{h}^{\mu\nu}(l)^*$. 

The mode function $f_k(x)$ thus has the behaviour at late times as: 
\begin{equation}
q_k(x)= e^{-ik.x}+ \int\frac{d^4l}{(2\pi)^3}\delta(l^2)k_{\mu}k_{\nu}\tilde{h}^{\mu\nu}(l)(k.l)^{-1}e^{-i(k-l).x}
\end{equation}	
The field expansion of the quantised scalar field $\Phi(x)$ at early times (which we denote as $\Phi^{(-)}(x)$) is exactly the same as that of $\Phi^{(0)}(x)$ since the background is flat. At late times, the background dynamics of the GW which has just passed through the observatory, changes the early time mode function $f_k(x)$ to $f_k^{(+)}(x)$ which has the functional form given above. Since $f_k$ and $f_k^*$ form a complete orthogonal basis, we can write: 
\begin{equation}
q_k(x)=\int d^3\vec{k'}\left(\alpha_{kk'}e^{-ik'.x}+\beta_{kk'}e^{ik'.x}\right) ~\label{qu}
\end{equation}
where $\alpha_{kk'}$ and $\beta_{kk'}$ are the Bogoliubov coefficients. Eqn. (\ref{qu}) can also be written as:
\begin{eqnarray}
&& q_k(x)=\int d^3\vec{k'}e^{-ik'.x}\delta^{(3)}(\vec{k}-\vec{k'})+ \nonumber \\
&& \int\frac{d^3\vec{k'}}{(2\pi)^32\omega_{\vec{k'}}}k_{\mu}k_{\nu}\tilde{h}^{\mu\nu}(k')(k.k')^{-1}e^{-i(k-k').x} ~\label{qu2}
\end{eqnarray}
where the new integration measure in the second term is as good as the old one. Comparing equations (\ref{qu}) and (\ref{qu2}), we get the Bogoliubov coefficients as: 
\begin{eqnarray}
\alpha_{kk'} &=& \delta^{(3)}(\vec{k}-\vec{k'}) ~\label{alph} \\
\beta_{kk'} &=& \frac{1}{(2\pi)^3}\frac{1}{2\omega_{\vec{k'}}}k_{\mu}k_{\nu}\tilde{h}^{\mu\nu}(k')(k.k')^{-1}e^{-ik.x} ~\label{bet}
\end{eqnarray}
Eqn. (\ref{alph}) is an expected result because of our perturbative treatment, and it is the only thing that remains when the background is flat, so that at initial and late times, we get $q_k=f_k$. Now as the mode function evolves from $f_k$, the vacuum state also changes due to the dynamic nature of the GW background. The scalar field expansion in the late time flat spacetime should be of the form:
\begin{equation}
\Phi^{(+)}(x)=\int \frac{d^3k}{(2\pi)^3 2\omega_{\vec{k}}} \left(q_k b_k + q_k^* b^\dagger_k\right)
\end{equation}
where $b_k|0\rangle_+=0$, $|0\rangle_+$ being the new vacuum state in the late time limit. Following Parker and Toms, we now introduce the covariant scalar product between two modes $f_1$ and $f_2$:
\begin{equation*}
(f_1,f_2)=i\int_{\Sigma}d^3v_\sigma\sqrt{-g}e^{\sigma}_{\mu}g^{\mu\nu}f_1\overset{\leftrightarrow}{\partial}_{\nu}f_2
\end{equation*}
where $e^{\sigma}_{\mu}$ is the unit normal to the space-like hypersurface $\Sigma$. So we have $g^{\mu\nu}e^{\sigma}_{\mu}e^{\sigma}_{\nu}=1$ and $g^{\mu\nu}e^{\sigma}_{\mu}a_{\nu}=0$ where $a_\mu\in\Sigma$. Upon choosing a frame where only the time component of the unit vector survives, and using the fact that $g_{\mu\nu}=\eta_{\mu\nu}+h_{\mu\nu}$, we can show that: 
\begin{eqnarray}
(f_1,f_2)=i\int d^3\vec{x}f_1\overset{\leftrightarrow}{\partial}_{0}f_2
\end{eqnarray}
We can also show that if $f_1$ and $f_2$ are solutions to the field equations in the curved background, this scalar product is independent of time. Using these facts, we get an expression for $b_k$ from equation (34) as: 
\begin{equation}
b_k=(q_k,\Phi^{(-)}(x))=\int d^3\vec{k'}\left(\alpha_{kk'}^*a_{k'}-\beta_{kk'}^*a_{k'}^{\dagger}\right)
\end{equation}
 We can now calculate the spectrum of created particles as the expectation value of the late time number operator 
$N^{(+)}_k=b_k^{\dagger}b_k$ with respect to the early time vacuum state $|0\rangle_-$: 
\begin{equation}
\left\langle N^{(+)}_{k}\right\rangle_-=_-\langle0|N^{(+)}_{k}|0\rangle_-=\int d^3\vec{k'}(2\pi)^3 2\omega_{\vec{k'}}|\beta_{kk'}|^2
\end{equation}
 Substituting the expression for $\beta_{kk'}$ from equation (5), we get the expression for the frequency spectrum of massless scalar particles produced, in terms of the gravitational wave amplitudes $\tilde{h}^{\mu\nu}$ as:
\begin{eqnarray}
\left\langle N^{(+)}_{k}\right\rangle_- &=& \int\frac{d^4k'}{(2\pi)^3}\delta(k'^2)k_{\mu}k_{\nu}k_{\alpha}k_{\beta}(k.k')^{-2} \cdot \nonumber \\
&& \cdot \tilde{h}^{\mu\nu}(k')\tilde{h}^{*\alpha\beta}(k') ~\label{enk}
\end{eqnarray}

The actual profile of the GWs detected at GW150914 and later runs of LVK should give us directly the amplitude of created massless scalars from the above expression. The details of our prediction for the spectrum of creation of scalar quanta will be treated in a subsequent paper. However, we can at once deduce the fact that the angular spectrum of the created particles is not spatially isotropic. This is because from equation (\ref{enk}), we see that $k.k'$ cannot vanish. But since $k$ and $k'$ are both non-parallel null vectors, we have that $k.k'=\omega_{\vec{k}}\omega_{\vec{k'}}(1-\cos\theta)$, where $\omega_k, \omega_{k'}$ are the respective frequencies. It follows that $\cos\theta\neq1$. This restriction stymies the possibility of an isotropic angular distribution.

\section{Outline of photon creation in a GW background}

Following our discussion in section 2, we use the projection operator written out in eqn. (\ref{vac}) and the physical electromagnetic vector potential in spacetime, obeying the transversality condition (\ref{trns}) as a {\it result} of such projection, not as a consequence of a choice. This implies that, the spacetime vector potential is given by  
\begin{equation}
A_{\mu}^{P}(x)=\sum_{\lambda=1}^{2}\int\frac{d^4k}{(2\pi)^3}\delta(k^2)\theta(k^0)\left\{a_{k}^{(\lambda)}\epsilon_{\mu}^{P(\lambda)}(k)e^{-ik.x}+h.c.\right\}
\end{equation}
where the physical (transverse) polarisation vectors satisfy the orthogonality relation:
\begin{eqnarray}
\epsilon^{\mu(\lambda)}_{P}(k)\bar{\epsilon}_{\mu}^{P(\lambda')}(k)=-\delta^{\lambda\lambda'}
\end{eqnarray}
In the GW background spacetime, the physical vector potential satisfies the EOM $\Box_gA_{\mu}^{P}=0$. Just as in the case of scalar fields, we perturbatively expand the vector potential as $A^{P}_{\mu}=A^{P(0)}_{\mu}+A^{P(1)}_{\mu}$ where $A^{P(0)}_{\mu}$ follows the flat spacetime KG equation and hence has the exactly identical field expansion as the above equation (2). At first order in perturbation theory we have: 
\begin{eqnarray}
\Box_{\eta}A_{\alpha}^{(1)} &=& \frac{1}{2}\left(\partial^{\nu}h_{\alpha}^{\sigma}+\partial_{\alpha}h^{\sigma\nu}-\partial^{\sigma}h^{\nu}_{\alpha}\right)\partial_{\nu}A^{(0)}_{\sigma}(x) \nonumber \\
&+& h^{\mu\nu}\partial_{\mu}\partial_{\nu}A^{(0)}_{\alpha}(x). \label{a1}
\end{eqnarray}

Proceeding as in the case of the scalar field in the earlier section, we obtain the perturbative solution of the quantum vector potential for photons as 
\begin{eqnarray}
&& A_{\alpha}(x) = \sum_r \int_k \delta(k^2) [\{ e^{-i k.x} \epsilon^{(r)}_{\alpha}(k) + \nonumber \\
&+& \int_l \delta(l^2) ( \frac14 {\tilde h}^{\beta}_{\alpha}(l) \epsilon^{(r)}_{\beta}(k) + \frac14 {\tilde h}^{\sigma}{\nu}(l) \epsilon^{(r)}_{\sigma}(k)k_{\nu} l_{\alpha} (k.l)^{-1} \nonumber \\
&-& \frac14 {\tilde h}^{\nu}_{\alpha}(l) k_{\nu} l^{\sigma} \epsilon^{(r)}_{\sigma}(k)(k.l)^{-1} + \frac12 {\tilde h}^{\mu \nu}(l) k_{\mu} k_{\nu} \epsilon^{(r)}_{\sigma} (k) (k.l)^{-1} ) \nonumber \\ 
&& e^{-i(k-l).x}  \}  a_k^{(r)} + h.c.], \label{qvp}
\end{eqnarray}
where, $\int_k \equiv \int (d^4k)/(2\pi)^3$ and similarly for $l$. Thus, the late time expansion of the photon field is
\begin{eqnarray}
A_{\mu}(x)= \sum_r \int_k \delta(k^2) \left[ a^{(r)}_k \epsilon^{(r)}_{\mu}(k) q_k + h.c.\right], \label{lata}
\end{eqnarray}
where the late time mode functions are given by
\begin{eqnarray}
q_k &=& \int d^3\vec{k'} [e^{-ik'.x}\delta^{(3)}(\vec{k}-\vec{k'})-\frac{1}{(2\pi)^{3}}\frac{1}{\omega_{\vec{k'}}} \{\frac{1}{8}\tilde{h}^{\sigma}_{\alpha}(k')P^{\alpha}_{\sigma}(k) \nonumber \\
&+& \frac{1}{8}\tilde{h}^{\sigma\nu}(k')k_{\nu}k^{'}_{\alpha}(k.k')^{-1} P_{\sigma}^{\alpha}(k) \nonumber \\
&-& \frac{1}{8}\tilde{h}^{\nu}_{\alpha}(k')k_{\nu}k^{'\sigma}(k.k')^{-1}P^{\alpha}_{\sigma}(k) \nonumber \\
&-&\frac{1}{2}\tilde{h}^{\mu\nu}(k')k_{\mu}k_{\nu}(k.k')^{-1} \} e^{-i(k-k').x} ]. \label{quk}
\end{eqnarray}
The mode evolution function is of the form:
\begin{eqnarray}
q_k=\int d^3\vec{k'}\left(\alpha_{kk'}e^{-ik'.x}+\beta_{kk'}e^{ik'.x}\right). \label{que}
\end{eqnarray}
Comparing eqn.s (\ref{quk}) and (\ref{que}), it is easy to read off the Bogoliubov coefficients $\alpha_{k k'}$ and $\beta_{kk'}$.

\section{Conclusion}
	
We have seen that we indeed get a non zero result for the amplitude of created particles. In this paper we have considered a massless scalar field and quantised it, and hence the particles produced are massless scalars. In principle we can take other fields with different spins and quantise them to get the production of all other kinds of particles because of the dynamical background. For example we can take the massless spin-1 field and the final result will be the production of electromagnetic radiation as a result of the creation of photons. Quantisation of Dirac field will result in production of fermions. This suggests that even in a binary black hole merger, the dynamical nature of the spacetime would result in particle creation which could be detected in principle. Thus we would expect the production of photons in a BBH merger even though there are no electromagnetic fields classically present. The question is whether such photons can be detected, if we can calculate their spectrum for the gravitational wave delected at LIGO. Due to the production of particles of various species by the gravitational wave background, even a binary black hole merger exhibits the phenomenon of Multi-Messenger Astronomy.

\vspace{10mm}

\section{Acknowledgements}

We profusely thank Professor Soumitra SenGupta for sharing his insights with us at the beginning of this work. We also thank Anarya Ray for many illuminating discussions. 
	
	%\nocite{*}
	%\bibliographystyle{plain}
	%\bibliography{bibliography}

\end{document}